\documentclass[12pt]{article}
\begin{document}
\title{$\overline{SL}(4,R)$ Embedding for a 3D\\ World Spinor Equation}

\author{Djordje \v Sija\v cki\thanks{email: sijacki@phy.bg.ac.yu} \\
Institute of Physics, P.O. Box 57, 11001 Belgrade, Serbia}

\date{}

\maketitle

\begin{abstract}
A generic-curved spacetime Dirac-like equation in $3D$ is constructed. It has,
owing to the $\overline{SL}(n,R)$ group deunitarizing automorphism, a
physically correct unitarity and flat spacetime particle properties.
The construction is achieved by embedding $\overline{SL}(3,R)$ vector
operator $X_{\mu}$, that plays a role of Dirac's $\gamma_{\mu}$ matrices, into
$\overline{SL}(4,R)$. Decomposition of the unitary irreducible spinorial
$\overline{SL}(4,R)$ representations gives rise to an explicit form of the
infinite $X_{\mu}$ matrices.
\end{abstract}

\section{Introduction}

The Dirac equation turned out to be one of the most successful theoretical
achievements of the 20th century physics. It describes the basic matter
constituents (both particles and fields), and very significantly, it played a
prominent role in ushering the minimal coupling and gauge principle in the
field of particle physics. The successes of the standard model based on this
equation established beyond doubt.

The objective of this work is to consider a Dirac-like equation describing a
spinorial field in a generic-curved spacetime. It turns out that there are
some quite nontrivial group theoretic features, related primarily to the
spinorial properties of the $SL(n,R)$ $\subset$ $Diff(n,R)$, $n\ge 3$ symmetry
groups, that make this quest difficult. For the sake of making a clear
parallel as well as understanding important distinctions between the group
theoretic structure of the flat and generic-curved spacetime spinorial field
equations, we recall briefly some relevant basic facts.

The Poincar\'e relativistic quantum field theory is a field theory obeying
Einstein's  principle of relativity, i.e. a field theory invariant w.r.t. the
Poincar\'e symmetry group (cf \cite{Ra, Rb}). The "representations on
states" are representations of the Poincar\'e group in the Hilbert space of
corresponding particle quanta. Unitarity of this Hilbert space is achieved by
making use of the unitary irreducible representations of the Poincar\'e group
$P(4) = T_4\wedge SO(1,3)$ ($T_4$ and $SO(1,3)$ being  the translational and
the Lorentz subgroups, respectively). Due to the Poincar\'e group semidirect
product nature, the particle states are characterized, besides the quantum
numbers of the representation invariants (mass and spin/helicity), by the
so-called little group quantum numbers (momentum and spin/helicity
projections). The representation space of the little group's part belonging to
the translational subgroup is infinite dimensional due to its noncompactnes
(continuous momentum values), while that of the little group's part
belonging to the Lorentz subgroup ($SO(3)$ for $m\ne 0$, i.e. $SO(2)$
$\subset$ $E(2)$ for $m = 0$) is finite dimensional. The "representations on
fields" are of the form 
\begin{eqnarray}
&&(D(a,\Lambda )\Phi_m)(x) = (D(\Lambda ))^n_m \Phi_n (\Lambda^{-1}(x-a)) \\
&&(a,\Lambda ) \in T_4\wedge SO(1,3) \nonumber ,
\end{eqnarray}
where $m,n$ enumerate a basis of the representation space of the field
components. The fact that finite-dimensional representations, $D(\Lambda )$,
of the Lorentz subgroup are, due to its noncompactnes, nonunitary is of no
physical relevance as long as one, by means of a field equation, prevents
propagations of all field components but those corresponding to a unitary
representation of the Lorentz subgroup part of the little group. In other
words, unitarity is imposed in the Hilbert space of the representations on
states only, while the field equations provide for a full Lorentz covariance
as well as restrict the field components in such a way that the physical
degrees of freedom are as given by the corresponding particle states.

The quantum-mechanical symmetry group $G_{q.m.} = \{ T(g) \}$ of a given
classical symmetry group $G = \{ g \}$, is given by a set of Hilbert space
operators satisfying $T(g_{1}) T(g_{2}) = e^{i \omega (g_{1},g_{2})}
T(g_{1}g_{2})$ in order to account for the physical Hilbert space ray
structure. The nontrivial solutions of $\omega (g_{1},g_{2})$ are obtained by
making use of the universal covering (topology features), and by finding
nonequivalent minimal extensions of the group $G$ by the $U(1)$ group (group
deformation features). In the case of a noncompact group, the topology
features are determined by its maximal compact subgroup. There are no
nontrivial deformations of the Poincar\'e group, however, there is a double
covering of it due to the fact that its maximal compact subgroup $SO(3)$
$\subset$ $SO(1,3)$ is double connected. The quantum-mechanical Poincar\'e
group $P(4)_{q.m.} = \overline{P}(4)$ is thus the double covering group of the
$P(4)$ group, i.e. $\overline{P}(4)/Z_{2} \simeq P(4)$, and the novel features
are the spinorial particles and fields. The double covering of the maximal
compact subgroup $SO(3)$ is the $Spin(3)$ group, that is isomorphic to
$SU(2)$, while the double covering of the Lorentz group $SO(1,3)$ is the
$Spin(1,3)$ group isomorphic to $SL(2,C)$ (for a detailed account of
group-theoretical structure of Poincar\'e spinors cf \cite{Rc}).

When passing from special to general relativity, one considers the group of
General Coordinate Transformations $GCT = Diff(4,R)$ instead of the Poincar\'e
group. The local (anholonomic) flat (tangent) spacetime tensors (Lorentz group
representations) are replaced by the world (holonomic) generic-curved
spacetime tensors (homogeneous $Diff(4,R)$ group representations). The
$Diff(4,R)$ tensor algebra is basically determined by the tensor calculus of
its linear subgroups $SL(4,R) \subset GL(4,R)$, i.e. for finite-component
tensors by the tensor calculus of  the corresponding compact groups $SU(4)
\subset U(4)$ (cf \cite{Rd, Re, Rf}). Note that linear, tensorial
representations of $Diff(4,R)$ $\simeq$ $\overline{Diff}(4,R)/Z_{2}$ can be
both finite-dimensional (nonunitary) and infinite-dimensional (unitary and
nonunitary).

The transition from special to general relativity, in the quantum case, is
achieved by considering the covering group $\overline{Diff}(4,R)$ of general
coordinate transformations instead of the covering of the Poincar\'e group.
This covering is again a double covering, since it is determined by the
covering of the $SO(4)$ maximal compact subgroup, which is in its turn a
double-connected Lie group: $\overline{SO}(4) = Spin(4)$, $SO(4) \simeq
Spin(4)/Z_{2}$. The same double covering result holds for any
dimension $D \geq 3$.

Let us consider now the question of the dimensionality of the fundamental
(lowest-dimensional) spinorial representations, i.e. the
dimensionality of the vector space in which the covering group is
defined. The defining space of the $SO(D-p, p)$ groups is $D$
dimensional, while the space of the corresponding covering
$Spin(D-p, p)$ groups is $2^{\left[ \frac{D-1}{2}\right]}$ dimensional. The
defining space dimensionality of the $Diff(D,R)$,
$\overline{Diff}(D,R)$ groups is given by the dimensionality of the
defining space of the $SL(D,R)$, $\overline{SL}(D,R)$ subgroups,
respectively. It turns out  that there are no finite-dimensional
spinorial representations of the $\overline{SL}(D,R)$ groups for $D
\ge 3$. The spinorial representations of these groups are
infinite dimensional, i.e. the $\overline{Diff}(D,R)$, $n \ge 
3$ groups are isomorphic to groups of infinite complex matrices
(cf \cite{R1}), and all their linear spinorial representations are
infinite dimensional as well.

It follows from the above considerations that a generalization of the Dirac
equation to a generic-curved  spacetime (not related to any $SO(m,n)$
orthogonal-type group) requires a knowledge of the infinite-component
spinorial fields, a construction of the corresponding vector operators
generalizing Dirac's $\gamma$ matrices, as well as physically satisfactory
unitarity properties defined by the appropriate particle states little
group unitary representations. One more comment is in order. Dirac's
equation was derived by factorizing the Poincar\'e second order invariant,
$(P^2 -m^2)\psi = 0$. Such an approach cannot be applied in our case
due to the fact that $P^2$ is not invariant w.r.t. nonorthogonal type
of groups, say $SL(D,R)$ etc.

The standard way of implementing spinors to General Relativity is to
consider either nonlinear spinorial $\overline{Diff}(4,R)$
representations w.r.t. its  $Spin(1,3)$ $\simeq$ $SL(2,C)$ subgroup, or, what
is nowadays customary, to make use of the tetrad formalism and spinorial
fields of a tangent flat spacetime. These spinors are spinors of the quantum
tangent-spacetime Lorentz group $Spin(1,3)$, however, they transforms as
scalars w.r.t. to the group of general coordinate transformations. In
contradistinction to the tensorial case, where there are both world and local
Lorentz fields that are mutually connected by appropriate tetrad-field
combinations, there exist only local Lorentz spinorial fields.

In this work we go beyond the Poincar\'e invariance and study a  Dirac-like
equation for infinite-component spinorial field that transforms w.r.t. linear
(single-valued) representations of the $\overline{Diff}(4,R)$ $\supset$
$\overline{GA}(4,R)$ $\supset$  $\overline{GL}(4,R)$ $\supset$
$\overline{SL}(4,R)$ $\supset$ $\overline{SO}(1,3)$ group chain. The Affine
$GA(4,R)$ group being a semidirect product, $GA(4,R)$ $=$ $T_4 \wedge
GL(4,R)$, of $4$-translations and the general linear group. In other words, we
consider a first-order wave equation for spinorial field, "world spinors"
\cite{R2, R3}, in a generic non-Riemannian spacetime of arbitrary torsion and
curvature. A flat spacetime version of this equation (constructed below),
i.e. the corresponding action, is that to be used in setting up a
metric-affine \cite{R4, R5} and/or affine \cite{R6} gauge theories of
gravitational interactions of spinorial matter.

Affine-invariant extensions of the Dirac equation have been considered
previously, however, lacking either required physical interpretation or
the actual invariance that goes beyond the Lorentz one. Mickelsson
\cite{R7} has constructed a truly $\overline{Diff}(4,R)$ $\supset$
$\overline{GL}(4,R)$ covariant extension of the Dirac equation,
however, its physical interpretation is rather unclear. The unitarity
problem as well as the questions of $\overline{GL}(4,R)$ irreducible
representations content and the physical particle states are not resolved.
Cant and Ne'eman \cite{R8} found a Dirac-type equation for
infinite-component fields of $\overline{SL}(4,R)$), however, this
equation does not stretch beyond Lorentz covariance. They use only a
subclass of $\overline{SL}(4,R)$ multiplicity-free representations
that does not allow for a $\overline{SL}(4,R)$ vector operator and an
extension to affine Dirac-like wave equation.

In a recent paper \cite{R9}, we considered a Dirac-type infinite-component
equation from the point of view of building it from physically well-defined
Lorentz subgroup components.

The aim of this paper is to provide an explicit construct of a world spinor
field equation that satisfies all conditions required by a correct
physical interpretation (unitarity, Poincar\'e particle interpretation for
the field components). Owing to the complexity of this task in the
$D=4$ case, as elaborated upon bellow, we confine in this paper to an
explicit construction in the $D=3$ case, where we make use of the
nonmultiplicity-free $\overline{SL}(3,R)$ representations, and
consequently achieve a full $\overline{Diff}(3,R)$ covariance. The
construction is achieved by embedding the relevant algebraic relations
into $D=4$, and by decomposing a physically motivated
$\overline{SL}(4,R)$ spinorial representation to $D=3$. All
expressions of section 2 are valid for any $D=n$, $n\geq 3$, thus we
write them in full generality. In the appendices we present, in an adjusted
notation, the group representation results that are essential for this
analysis.

\section{World spinors}

The finite-dimensional world tensor fields in $R^{n}$ are
characterized by the nonunitary irreducible representations of the
general linear subgroup $GL(n,R)$ of the Diffeomorphism group
$Diff(n,R)$. In the flat-space limit these representations split up
into $SO(1,n-1)$  ($SL(2,C)/Z_2$ for $n=4$) irreducible pieces. The
corresponding particle states are defined in the tangent flat-space
only. They are characterized by the unitary irreducible
representations of the (inhomogeneous) Poincar\'e group $P(n) =
T_n\wedge SO(1,n-1)$, and their components are enumerated by the
"little" group unitary representations (e.g. $T_{n-1}\otimes SO(n-1)$
for $m\ne 0$). In the generalization to world spinors, the double
covering $\overline{SO}(1,n-1)$ of the $SO(1,n-1)$ group, that
characterizes a Dirac-type fields in $D=n$ dimensions, is enlarged to
the $\overline{SL}(n,R) \subset \overline{GL}(n,R)$ group, while
$SA(n,R) = T_n\wedge \overline{SL}(n,R)$ is to replace the Poincar\'e
group itself. Affine "particles" are now characterized by the unitary
irreducible representations of the $\overline{SA}(n,R)$ group, i.e. by
the nonlinear unitary representations over an appropriate "little"
group (e.g. $T_{n-1}\otimes \overline{SL}(n-1,R) \supset
T_{n-1}\otimes \overline{SO}(n-1)$, for $m\ne 0$).

A mutual particle-field correspondence is achieved by requiring (i)
that fields have appropriate mass (Klein-Gordon-like equation
condition), and (ii) that  the subgroup of the field-defining
homogeneous group, that is isomorphic to the homogeneous part of the
"little" group, is represented unitarily. Furthermore, one has to
project away all little group representations except the one that
characterizes the (physical, i.e. propagating) particle states.

A physically correct picture, in the affine case, is obtained by
making use of the $\overline{SA}(n,R)$ group unitary (irreducible)
representations for "affine" particles.  The affine-particle states
itself are characterized by the unitary (irreducible) representations
of the $T_{n-1}\otimes \overline{SL}(n-1,R)$ "little" group. The
"intrinsic" part of these representations is necessarily
infinite dimensional due to noncompactness of the $SL(n,R)$ group.
The corresponding affine fields are described by nonunitary
infinite-dimensional $\overline{SL}(n,R)$ representations, that should
be unitary when restricted to $\overline{SL}(n-1,R)$, the homogeneous
part of the "little" subgroup. Therefore, the first step towards world
spinor fields is a construction of infinite-dimensional nonunitary
$\overline{SL}(n,R)$ representations, that are unitary when restricted
to the $\overline{SL}(n-1,R)$ subgroup. Each of these fields reduce to
an infinite sum of (nonunitary) finite-dimensional
$\overline{SO}(1,n-1)$ fields having the usual relativistic field
interpretation.

{\it The deunitarizing automorphism}. The unitarity properties, that
ensure correct physical interpretation of the affine fields, can be
achieved by combining the unitary (irreducible) representations with
the so-called "deunitarizing" automorphism of the
$\overline{SL}(n,R)$ group \cite{R1}.

The commutation relations of the $\overline{SL}(n,R)$ generators
$Q_{ab}$, $a,b$ $=$ \hfill\break $0, 1, \dots , n-1$ are
\begin{equation}
[Q_{ab},Q_{cd}] = i(\eta_{bc}Q_{ad} - \eta_{ad}Q_{cb}) ,
\end{equation}
where $\eta_{ab} = diag(+1, -1, \dots , -1)$. The important
subalgebras are as follows:

(i) $so(1,n-1)$: The $M_{ab} = Q_{[ab]}$  operators generate the
Lorentz-like subgroup $\overline{SO}(1,n-1)$ with $J_{ij}$ (angular
momentum) and $K_{i} = M_{0i}$ (the boosts) $i,j = 1,2,\dots ,n-1$.

(ii) $so(n)$: The $J_{ij}$ and $N_{i} = Q_{\{0i\}}$ operators generate
the maximal compact subgroup $\overline{SO}(n)$.

(iii) $sl(n-1)$: The $J_{ij}$ and $T_{ij} = Q_{\{ij\}}$ operators
generate the $\overline{SL}(n-1,R)$ subgroup - the "little" group of
the massive particle states.

The $\overline{SL}(n,R)$ commutation relations are invariant under
the automorphism,
\begin{eqnarray}
 &J^\prime_{ij} = J_{ij}\ ,\quad K^\prime_{i} = iN_{i}\ ,\quad
N_{i}^\prime = iK_{i}\ , \nonumber \\
 &T^\prime_{ij} = T_{ij}\ ,\quad
T^\prime_{00} = T_{00} \ (= Q_{00})\ ,
\end{eqnarray}
so that $(J_{ij},\ iK_{i})$ generate the ``new'' compact
$\overline{SO}(n)^\prime$ group, while $(J_{ij},\ iN_{i})$ generate
the ``new'' noncompact $\overline{SO}(1,n-1)^\prime$ group.

For the spinorial particle states we start with the basis
vectors of the unitary irreducible representations of
$\overline{SL}(n,R)^\prime$, so that the compact subgroup finite
multiplets correspond to $\overline{SO}(n)^\prime$: $(J_{ij},\
iK_{i})$ while $\overline{SO}(1,n-1)^\prime$: $(J_{ij},\ iN_{i})$ is
represented by unitary infinite-dimensional representations. We now
perform the inverse transformation and return to the unprimed
$\overline{SL}(n,R)$ for our physical identification:
$\overline{SL}(n,R)$ is represented nonunitarily, the compact
$\overline{SO}(n)$ is represented by nonunitary infinite
representations while the Lorentz group is represented by nonunitary 
finite representations. These finite-dimensional nonunitary Lorentz
group representations are necessary in order to ensure a correct
particle interpretation (i.e. boosted proton remain proton). Note that
$\overline{SL}(n-1,R)$, the stability subgroup of
$\overline{SA}(n,R)$, is invariant w.r.t. the deunitarizing
automorphism, and thus it remains represented unitarily.

The world spinor fields transform w.r.t. $\overline{Diff}(n,R)$ as
follows
\begin{eqnarray}
&&(D(a,\bar f)\Psi_M) (x)
= (D_{ \overline{Diff}_0(n,R)}(\bar f))^N_M \Psi_N
(f^{-1}(x-a)), \\
&&(a,\bar f) \in T_{n} \wedge \overline{Diff}_0(n,R), \nonumber
\end{eqnarray}
\noindent where $\overline{Diff}_0(n,R)$ is the homogeneous part
of $\overline{Diff}(n,R)$, while $f$ is the element corresponding to
$\bar f$ in $Diff(n,R)$. The $D_{\overline{Diff}_0(n,R)}$
representations can be reduced to direct sum of infinite-dimensional
$\overline{SL}(n,R)$ representations.  As a matter of fact, we
consider here those representations of
$\overline{Diff}_0(n,R)$ that are nonlinearly realized over the
maximal linear subgroup $\overline{SL}(n,R)$.

The affine "particle" states transform in the following way:
\begin{equation}
D(a, \bar s) \rightarrow
e^{i a\cdot (sp)} D_{\overline{SL}(n,R)} (L^{-1}(sp)\bar s
L(p)) ,\quad (a,\bar s) \in T_n\wedge \overline{SL}(n,R),
\end{equation}
where $L \in \overline{SL}(n,R)/\overline{SL}(n-1,R)$, and $p$ is
the $n$-momentum label. The unitarity properties of various
representations in this expression are as described above.

Provided the relevant $\overline{SL}(n,R)$ representations are
known, one can first define the corresponding general/special
Affine spinor fields, $\Psi_{A}(x)$, in the tangent to $R^{n}$, and
than make use of the infinite-component pseudo-frame fields
$E^{A}_{M}(x)$, "alephzeroads", that generalize the tetrad fields of
$R^{4}$ \cite{R2}. Let us define a pseudo-frame $E^{A}_{M}(x)$ s.t.
\begin{equation}
\Psi_{M}(x) = E^{A}_{M}(x) \Psi_{A}(x),
\end{equation}
where $\Psi_{M}(x)$ and $\Psi_{A}(x)$ are the world (holonomic)
and local affine (anholonomic) spinor fields, respectively. The
$E^{A}_{M}(x)$ (and their inverses $E^{M}_{A}(x)$) are thus
infinite matrices related to the quotient
$\overline{Diff}_{0}(n,R)/\overline{SL}(n,R)$. Their
infinitesimal transformations are
\begin{equation}
\delta E^{A}_{M}(x) = i\epsilon^{a}_{b}(x) \{Q_{a}^{b}\}^{A}_{B}
E^{B}_{M}(x) +
\partial_\mu \xi^{\nu} e^{a}_{\nu} e^{\mu}_{b}\{Q^{a}_{b}\}^{A}_{B}
 E^{B}_{M}(x),
\end{equation}
where  $\epsilon^{a}_{b}$ and $\xi^\mu$ are group parameters of
$\overline{SL}(n,R)$ and \hfill\break
$\overline{Diff}(n,R)/\overline{Diff}_{0}(n,R)$ respectively,
while $e^{a}_{\nu}$ are the standard $n$-bine frame fields.

The transformation properties of the world spinor fields
themselves are given as follows:
\begin{equation}
\delta \Psi^{M}(x) = i\big\{ \epsilon^{a}_{b}(x) E^{M}_{A}(x)
\big(Q_{a}^{b}\big)^{A}_{B} E^{B}_{N}(x) + \xi^{\mu} \big[
\delta^{M}_{N}\partial_{\mu} +
E^{M}_{B}(x)\partial_{\mu}E^{B}_{N}(x)\big] \big\} \Psi^{N}(x).
\end{equation}
The $\big( Q^{b}_{a} \big)^{M}_{N} = E^{M}_{A}(x)
\big(Q_{a}^{b}\big)^{A}_{B} E^{B}_{N}(x)$ is the holonomic form of
the $\overline{SL}(n,R)$ generators given in terms of the
corresponding anholonomic ones. The $\big( Q^{b}_{a} \big)^{M}_{N}$
and  $\big(Q_{a}^{b}\big)^{A}_{B}$ act in the spaces of spinor fields
$\Psi_{M}(x)$ and $\Psi_{A}(x)$, respectively.

The above outlined construction allows one to define a fully
$\overline{Diff}(n,R)$ covariant Dirac-like wave equation for the
corresponding world spinor fields provided a Dirac-like wave
equation for the $\overline{SL}(n,R)$ group is known. In other
words, one can lift up an $\overline{SL}(n,R)$ covariant equation of
the form
\begin{equation}
\big( ie^{\mu}_{a} \big(X^{a}\big)^{B}_{A} \partial_{\mu} - M \big)
\Psi_{B}(x) = 0,
\end{equation}
to a $\overline{Diff}(n,R)$ covariant equation
\begin{equation}
\big( ie^{\mu}_{a} E^{A}_{M} \big(X^{a}\big)^{B}_{A}E^{N}_{B}
\partial_{\mu} - M \big) \Psi_{N}(x) = 0,
\end{equation}
where the former equation exists provided a spinorial
$\overline{SL}(n,R)$ representation for $\Psi$ is given, such that the
corresponding representation Hilbert space is invariant w.r.t. $X^{a}$
action. Thus, the crucial step towards a Dirac-like world spinor
equation is a construction of the vector operator $X^{a}$ in the space
of $\overline{SL}(n,R)$ spinorial representations.

In the above considerations, we treated curved spacetime as externally
defined, and were interested in describing a propagation of a spinorial
field $\Psi (x)$ in such a background. One can now pose a question of
coupling the matter described by our infinite-component spinorial
field to the interactions of both gravitational and nongravitational
type. The standard way is provided by localizing the relevant global
symmetry groups. The full procedure would be to start by requiring local
invariance of an action given in terms of globally covariant fields,
and to derive the corresponding interacting field equations with
ordinary derivatives replaced by the appropriate covariant ones. Let
the global symmetry group be given by $G_{grav}\otimes G_{int}$,
where say $G_{grav} = \overline{SA}(n,R)$, and $G_{int} =
SU(3)_{c}\otimes SU(2)_{L}\otimes SU(1)$ of the standard
model. Localization of this symmetry provides for the gauge potential
fields, and the covariant derivative form. In our case, the spinorial
equation that describes propagation and appropriate couplings reads
symbolically
\begin{equation}
\Big( i \big(X^{a}\big)^{B}_{A}
\big( e^{\mu}_{a}\partial_{\mu}
-i Q^{b}_{a}\Gamma^{a}_{\ b\mu}
-ig_{int} \lambda_{k}A^{k}_{\mu} \big) -M \Big)
\Psi_{B}(x) = 0 .
\end{equation}
The gravitational gauge potentials are $e^{a}_{\mu}$ and
$\Gamma^{a}_{\ b\mu}$, $\lambda_{k}$ stand for all internal
group generators, while the corresponding gauge potentials and
coupling constants are $A^{k}_{\mu}$ and $g_{int}$ respectively.

\section{$SL(3,R)$ vector operator}

In order to illustrate the difficulties one encounters when considering 
action of an $SL(3,R)$ vector operator $X$ acting in the Hilbert space
of field components, i.e. when coupling the vector representation and the
representation of some field $\Psi$, let us consider at first the case
of finite-dimensional field representations (we leave aside additional
difficulties related to the question of auxiliary fields etc).

Let us consider a field $\Psi$ transforming w.r.t. a single
irreducible representation $D_{\Psi}(g)$.  Algebraic condition
necessary for a wave equation construction is that in the reduction of
the product of the vector representation $D_{X}(g)$ and the
representation $D_{\Psi}(g)$ one finds either the representation
$D_{\Psi}(g)$ itself or its contragradient representation
$D^{T}_{\Psi}(g^{-1})$, thus resulting in a use of $D_{\Psi}(g)$ or
$D_{\Psi}(g)\oplus D^{T}_{\Psi}(g^{-1})$, respectively. In
contradistinction to the  $3$-dimensional Lorentz group $SO(1,2)$
case, where this condition is always satisfied due to $D^{(1)} \otimes
D^{(j)} = D^{(j-1)} \oplus D^{(j)} \oplus D^{(j+2)}$ ($j$ being the
$\overline{SO}(1,2)$ representation label: $0$, $\frac{1}{2}$, $1$,
...), the $SL(3,R)$ case is more complex. Let us denote an $SL(3,R)$
irreducible representations by a Young tableau $[p, q]$, where $p$ and
$q$ are respectively, the number of boxes in the first and the second
row. The coupling of the vector representation $[1,0]$ with a generic
irreducible representation $[p,q]$, gives in general $[1,0] \otimes
[p,q]$ $=$ $[p+1,q] \oplus [p,q+1] \oplus [p-1,q-1]$. Taking into
account that the contragradient representation to the representation
$[p,q]$ is given by $[p,p-q]$, we find that one can satisfy the above
necessary algebraic condition only in the special case of a reducible
$SL(3,R)$ representation $[2q+1,q] \oplus [2q+1,q+1]$.

It is well known that one can indeed satisfy the commutation relations
\begin{equation}
[M_{ab}, X_{c}] = i(\eta_{ac}X_b - \eta_{bc}X_a).
\end{equation}
in the Hilbert space of any $SO(1,2)$ irreducible representation. 
However, in order for an $SO(1,2)$ vector to be an $SL(3,R)$
vector as well, it has to satisfy additionally the following
commutation relations
\begin{equation}
[T_{ab}, X_{c}] = i(\eta_{ac}X_b + \eta_{bc}X_a) .
\end{equation}
This is a much harder task to achieve, and in principle, one can find
nontrivial solutions only for particular representation spaces.

Let us turn now to infinite-dimensional representations.
The multiplicity free (ladder) unitary (infinite-dimensional)
irreducible representations $D^{ladd)}_{SL(3,R)}(0)$, and
$D^{ladd)}_{SL(3,R)}(1)$, with the $SO(3)$ subgroup content given by
$\{ j\}$  $=$ $\{ 0,2,4, \dots \}$, and $\{ j\}$ $=$ $\{ 1,3,5, \dots \}$
respectively, can be viewed as the limiting cases of the series of
finite-dimensional representations $[0,0]$, $[2,0]$, $[4,0]$, ..., and
$[1,0]$, $[3,0]$, $[5,0]$, ..., respectively. Upon the coupling with
the $SL(3,R)$ vector representation $[1,0]$, one has $[1,0]\otimes
[2n,0] \supset [2n+1,0]$, and $[1,0]\otimes [2n+1,0] \supset
[2n+2,0]$, ($n=0,1,2,\dots$). It would seem possible, at the first
site, to represent the vector operator $X$ in the Hilbert space of the
$D^{ladd)}_{SL(3,R)}(0) \oplus D^{ladd)}_{SL(3,R)}(1)$ representation.
However, the resulting representations obtained after the $X$ action
have different values of the Casimir operators and thus define new
(mutually  orthogonal) Hilbert spaces. 

If one starts, for instance, with the representation space of the
scalar representation $[0,0]$, the vector operator action would
produce the space of the vector representation $[1,0]$ itself, in the
next act one gets the spaces of the representations $[2,0]$ and
$[1,1]$, and so on. Therefore, unless some additional algebraic
constraints are imposed, one would end up (independently of the
starting representation) with an infinite-dimensional space consisting
of the representation spaces of all $SL(3,R)$ (nonunitary) irreducible
representations.

A rather efficient way to impose additional algebraic constraints on
the vector operator $X$ consists in embedding it into a non-Abelian
Lie-algebraic structure. The minimal semi-simple Lie algebra that
contains both the $SL(3,R)$ algebra and the corresponding vector
operator $X$ is the $SL(4,R)$ algebra. There are two $SL(3,R)$ vector
operators: $A = (A^a , a=1,2,3)$ and $B = (B_a , a=1,2,3)$, in the
$SL(4,R)$ algebra that transform w.r.t. $[1,0]$ and $[1,1]$ $SL(3,R)$
representations, respectively. Components of each of them mutually
commute, while their commutator yields the $SL(3,R)$ generators
themselves, i.e. 
\begin{equation}
[A^{a}, A^{b}] = 0,\quad [B_{a}, B_{b}] = 0,\quad [A^{a}, B_{b}] =
i Q^{a}_{b}.
\end{equation}
Now, due to the $SL(4,R)$ algebra constraints, any irreducible
representation (or an arbitrary combination of them) of $SL(4,R)$
defines a Hilbert space that is invariant under the action of an
$SL(3,R)$ vector operator proportional to $A$ or $B$. As an
example, let as consider a ten-dimensional $SL(4,R)$
representation given by the Young tableau $[2,0,0]$. This
representation reduces to $[2,0]$, $[1,0]$ and $[0,0]$
representations of $SL(3,R)$ that are of dimension $6$, $3$, and
$1$, respectively. The action of the $SL(3,R)$ $3$-vector $X$ is as
follows: 
\begin{equation}
X:\quad \left\{ \begin{array}{l} {[0,0] \rightarrow [1,0]} \\
{[1,0] \rightarrow [2,0]}
\\ {[2,0] \rightarrow 0} \end{array} \right.
\end{equation}
where $X:\ [2,0] \to 0$ is due to constraints enforced by the
$SL(4,R)$ algebra. In the reduction of the ten-dimensional space
to $6+3+1$, the vector operator $X$ has the following block-matrix
form:
\begin{equation}
X \sim \left(
\begin{array}{ccc}
0_{(6\times 6)} & a_{(6\times 3)} & 0_{(6\times 1)} \\ b_{(3\times
6)} & 0_{(3\times 3)} & a_{(3\times 1)} \\ 0_{(1\times 6)} &
b_{(1\times 3)} & 0_{(1\times 1)} \\
\end{array}
\right)
\end{equation}
where the nonzero matrix elements $a_{(m\times n)}$ and
$b_{(m\times n)}$ correspond to the $SL(4,R)$ generators $A$ and
$B$, respectively.

Let us consider now the $\overline{SL}(3,R)$ spinorial
representations, that are necessarily infinite-dimensional. There
is a unique multiplicity-free ("ladder") unitary irreducible
representation of the $\overline{SL}(3,R)$ group,
$D^{(ladd)}_{\overline{SL}(3,R)}(\frac{1}{2})$, that in the
reduction w.r.t. its maximal compact subgroup $\overline{SO}(3)$
yields,
\begin{equation}
D^{(ladd)}_{\overline{SL}(3,R)}(\frac{1}{2}) \supset
D^{(\frac{1}{2})}_{\overline{SO}(3)} \oplus
D^{(\frac{5}{2})}_{\overline{SO}(3)} \oplus
D^{(\frac{9}{2})}_{\overline{SO}(3)} \oplus \dots
\end{equation}
i.e. it has the following $J$ content: $\{J\}$ $=$
$\{\frac{1}{2},\ \frac{5}{2},\ \frac{9}{2},\ \dots\}$.

Owing to the fact that the $\overline{SO}(3)$ and/or
$\overline{SL}(3,R)$ vector operator can have nontrivial matrix
elements only between the $\overline{SO}(3)$ states such that
$\Delta J = 0, \pm 1$, it is obvious (on account of the
Wigner-Eckart theorem) that all $X$-operator matrix elements in
the Hilbert space of the
$D^{(ladd)}_{\overline{SL}(3,R)}(\frac{1}{2})$ representation
vanish. The same holds for the two classes of tensorial ladder
unitary irreducible representations
$D^{(ladd)}_{\overline{SL}(3,R)}(0;\sigma_2)$ and
$D^{(ladd)}_{\overline{SL}(3,R)}(1;\sigma_2)$, $\sigma_2 \in R$,
with the $J$ content $\{ J \} = \{ 0,2,4,\dots \}$ and $\{ J \} =
\{ 1, 3, 5,\dots \}$.

Let us consider now the case of $\overline{SL}(3,R)$ unitary irreducible
representations with nontrivial multiplicity w.r.t. its maximal compact
subgroup $\overline{SO}(3)$. An efficient way to construct these
representations explicitly is to set up a Hilbert space of
square-integrable functions $H = L^2([\overline{SO}(3) \otimes
\overline{SO}(3)]^d, \kappa )$, over the diagonal subgroup of the two
copies of the $\overline{SO}(3)$ subgroup, with the group action to the
right defining the group/representation itself while the group action to
the left accounts for the  multiplicity. Here, $\kappa$ denotes a
kernel of a Hilbert space scalar product, that is generally more
singular than the Dirac delta function in order to account for all
types of $\overline{SL}(3,R)$ unitary irreducible representations. Let
us make use of the canonical (spherical) basis in this space,
i.e. $\sqrt{2J+1} D^{J}_{K M} (\alpha , \beta , \gamma )$, where $J$
and $M$ are the representation labels defined by the subgroup chain
$\overline{SO}(3) \supset \overline{SO}(2)$, while $K$ is the label of
the extra copy $\overline{SO}(2)_L \subset \overline{SO}(3)_L$ that
describes nontrivial multiplicity. Here, $-J \le K,M \le +J$, and for
each allowed $K$ one has $J \ge K$, i.e. $J = K, K+1, K+2, \dots$.

A generic $3$-vector operator ($J=1$) in the spherical basis ($\alpha
= 0, \pm 1$) reads:
\begin{equation}
X_{\alpha} = {\cal X}_{(0)} D^{(1)}_{0 \alpha} (k) + {\cal
X}_{(\pm 1)} [D^{(1)}_{+1\alpha}(k) + D^{(1)}_{-1\alpha}(k)],
\quad k \in \overline{SO}(3) .
\end{equation}
The corresponding matrix elements between the states of two
unitary irreducible $\overline{SL}(3,R)$ representations that are
characterized by the labels $\sigma$ and $\delta$ are given as
follows:
\begin{eqnarray}
&&\left< \begin{array}{c}(\sigma '\ \delta ') \\ J'
\\ K'\ M'\end{array} \right| X_{\alpha} \left|
\begin{array}{c}(\sigma\ \delta ) \\ J \\ K\ M \end{array}
\right> = (-)^{J'-K'} (-)^{J'-M'} \sqrt{(2J'+1)(2J+1)} \nonumber \\
&&\times\left(
\begin{array}{ccc}\ J' & 1 & J \\ -M' & \alpha & M \\ \end{array}
\right) \Bigg\{
 {{\cal X}_{(0)}}^{(\sigma ' \delta ' \sigma \delta )}_{J'J}
\left(\begin{array}{ccc}\ J' & 1 & J \\ -K' & 0 & K
\end{array} \right) \nonumber \\
&&+ {{\cal X}_{(\pm 1)}}^{(\sigma ' \delta ' \sigma \delta
)}_{J'J} \left[ \left( \begin{array}{ccc}\ J' & 1 & J \\ -K' & 1 &
K \end{array} \right) + \left( \begin{array}{ccc}\ J' & \ 1 & J \\
-K' & -1 & K \end{array} \right) \right]\Bigg\}.
\end{eqnarray}
Therefore, the action of a generic $\overline{SL}(3,R)$ vector
operator on the Hilbert space of some nontrivial-multiplicity
unitary irreducible representation produces the $\Delta J = 0, \pm
1$, as well as the $\Delta K = 0, \pm 1$ transitions. Owing to the
fact that the states of a unitary irreducible $\overline{SL}(3,R)$
representation are characterized by the $\Delta K = 0, \pm 2$
condition, it is clear that the $\Delta K = \pm 1$ transitions due
to $3$-vector $X$ take place between the states of mutually
inequivalent $\overline{SL}(3,R)$ representations whose
multiplicity is characterized by the $K$ values of opposite
evenness. In analogy to the finite-dimensional (tensorial)
representation case, the repeated applications of a vector
operator on a given unitary irreducible (spinorial and/or
tensorial) $\overline{SL}(3,R)$ representation would yield, a
priori, an infinite set of irreducible representations. Due to an
increased mathematical complexity in the case of infinite-dimensional
representations, some additional algebraic constraints imposed on
the vector operator $X$ would be even more desirable than in the
finite-dimensional case. The most natural option is to embed the
$\overline{SL}(3,R)$ $3$-vector $X$ together with the
$\overline{SL}(3,R)$ algebra itself into the (simple) Lie algebra
of the $\overline{SL}(4,R)$ group. Any spinorial (and/or
tensorial) $\overline{SL}(4,R)$ unitary irreducible representation
provides a Hilbert space that can be decomposed w.r.t.
$\overline{SL}(3,R)$ subgroup representations, and most
importantly that is invariant under the action of the vector
operator $X$. Moreover, an explicit construction of the starting
$\overline{SL}(4,R)$ representation would provide additionally for an
explicit form of $X$.

\section{Embedding into $\overline{SL}(4,R)$}

The $\overline{SL}(4,R)$ group, the double covering of the $SL(4,R)$
group, is a $15$-parameter non-compact Lie group, whose defining
(spinorial) representation is given in terms of infinite
matrices. All spinorial (unitary and nonunitary) representations of
$\overline{SL}(4,R)$ are necessarily infinite-dimensional; the
finite-dimensional tensorial representations are nonunitary, while the
unitary tensorial representations are infinite-dimensional. The
$\overline{SL}(4,R)$ commutation relations in the Minkowski space are
given by,
\begin{equation} 
[Q_{ab}, Q_{cd}] = i\eta_{bc} Q_{ad} - i\eta_{ad} Q_{cb}.
\end{equation}
where, $a,b,c,d = 0,1,2,3$, and $\eta_{ab} = diag(+1, -1, -1, -1)$, while
in the Euclidean space they read,
\begin{equation} 
[Q_{ab}, Q_{cd}] = i\delta_{bc} Q_{ad} - i\delta_{ad} Q_{cb}.
\end{equation}
where, $a,b,c,d = 1,2,3,4$, and $\delta_{ab} = diag(+1, +1, +1, +1)$

The relevant subgroup chain reads
\begin{equation}
\begin{array}{ccc}
\overline{SL}(4,R) & \supset & \overline{SL}(3,R) \\ \cup &  & \cup \\
\overline{SO}(4), \overline{SO}(1,3) & \supset & \overline{SO}(3),
\overline{SO}(1,2).
\end{array}
\end{equation}
We denote by $R_{mn},\ (m,n=1,2,3,4)$ the $6$ compact generators of the
maximal compact subgroup $\overline{SO}(4)$ of the $\overline{SL}(4,R)$
group, and the remaining $9$ noncompact generators (of the
$\overline{SL}(4,R)/\overline{SO}(4)$ coset) by $Z_{mn}$.

In the $\overline{SO}(4) \simeq SU(2)\otimes SU(2)$ spherical basis, the
compact operators are $J^{(1)}_i = \frac{1}{2}(\epsilon_{ijk} R_{jk} +
R_{i4})$ and $J^{(2)}_i = \frac{1}{2}(\epsilon_{ijk} R_{jk} - R_{i4})$,
while the noncompact generators we denote by $Z_{\alpha\beta},\ (\alpha
,\beta = 0, \pm 1)$, and they transform as a $(1, 1)$-tensor operator
w.r.t. $SU(2)\otimes SU(2)$ group. The minimal set of commutation
relations in the spherical basis reads
\begin{eqnarray}
&&[J^{(p)}_0, J^{(q)}_{\pm}] = \pm\delta_{pq} J^{(p)}_{\pm}, \quad
[J^{(p)}_+, J^{(q)}_-] = 2\delta_{pq} J^{(p)}_0, \quad (p,q=1,2),
\nonumber \\
&&[J^{(1)}_0, Z_{\alpha\beta}] = \alpha Z_{\alpha\beta}, \quad
[J^{(1)}_{\pm}, Z_{\alpha\beta}] = \sqrt{2 - \alpha (\alpha \pm 1)}
Z_{\alpha \pm 1\ \beta} \nonumber \\ 
&&[J^{(2)}_0, Z_{\alpha\beta}] = \beta
Z_{\alpha\beta}, \quad [J^{(2)}_{\pm}, Z_{\alpha\beta}] = \sqrt{2 - 
\beta (\beta \pm 1)} Z_{\alpha \beta\pm 1} \nonumber \\ 
&& [Z_{+1\ +1}, Z_{-1\ -1}] = - (J^{(1)} + J^{(2)}).
\end{eqnarray}
The $\overline{SO}(3)$ generators are $J_i = \epsilon_{ijk}J_{jk},\
J_{ij}\equiv R_{ij},\  (i,j,k =1,2,3)$, while the traceless
$T_{ij}=Z_{ij}\ (i,j=1,2,3)$ define the coset
$\overline{SL}(3,R)/\overline{SO}(3)$. In the $\overline{SO}(3)$
spherical basis the compact operators are $J_0,\ J_{\pm 1}$, while the
noncompact ones $T_{\rho}, \ (\rho = 0, \pm 1, \pm 2)$ transform w.r.t.
$\overline{SO}(3)$ as a quadrupole operator. The corresponding minimal
set of commutation relations reads
\begin{eqnarray}
&&[J_0, J_{\pm}] = \pm J_{\pm}, \quad [J_+, J_- ] = 2J_0 \nonumber \\ 
&& [T_{+2}, T_{-2}] = - 4J_0.
\end{eqnarray}

There are three (independent) $\overline{SO}(3)$ vectors in the algebra
of the $\overline{SL}(4,R)$ group. They are: (i) the $\overline{SO}(3)$
generators themselves, (ii) $N_i \equiv R_{i4} = Q_{i0} + Q_{0i}$, and
(iii) $K_i \equiv Z_{i4} = Q_{i0} - Q_{0i}$. From the latter two, one
can form the following linear combinations:
\begin{equation}
A_i = \frac{1}{2}(N_i + K_i) = Q_{i0}, \quad B_i = \frac{1}{2}(N_i - K_i)
= Q_{0i}.
\end{equation}
The commutation relations between $N$, $K$, $A$, and $B$ and the
$\overline{SL}(3,R)$ generators read
\begin{eqnarray}
&&{}[J_i, N_j] = i\epsilon_{ijk}N_k, \quad [T_{ij}, N_k] =
i(\delta_{ik}K_j + \delta_{jk}K_i), \nonumber \\ 
&&{}[J_i, K_j] = i\epsilon_{ijk}K_k, \quad [T_{ij}, K_k] = 
i(\delta_{ik}N_j + \delta_{jk}N_i), \nonumber \\ 
&&{}[J_i, A_j] = i\epsilon_{ijk}A_k, \quad [T_{ij},
A_k] = i(\delta_{ik}A_j + \delta_{jk}A_i), \nonumber \\ 
&&{}[J_i, B_j] = i\epsilon_{ijk}B_k, \quad [T_{ij}, B_k] = 
-i(\delta_{ik}B_j + \delta_{jk}B_i).
\end{eqnarray}
It is clear from these expressions that only $A_i$ and $B_i$ are
$\overline{SL}(3,R)$ vectors as well. More precisely, $A$ transforms
w.r.t. $\overline{SL}(3,R)$ as the three-dimensional representation
$[1,0]$, while $B$ transforms as its contragradient three-dimensional
representation $[1,1]$.

To summarize, either one of the two nontrivial $\overline{SL}(3,R)$
vector choices 
\begin{equation}
X_i \sim A_i, \quad X_i \sim B_i
\end{equation}
insures that a Dirac-like wave equation $(iX\partial - m)\Psi (x) = 0$
for a (infinite-component) spinorial field is fully $\overline{SL}(3,R)$
covariant. The choices
\begin{equation}
X_i \sim N_i, \quad X_i \sim K_i ,
\end{equation}
would yield wave equations that are Lorentz covariant only, even
though the complete $\overline{SL}(3,R)$ acts invariantly in the space
of $\Psi (x)$ components. It goes without saying that the correct
unitarity properties can be accounted for by making use of the
deunitarizing automorphism, as discussed above.

\section{Reduction of the $\overline{SL}(4,R)$ multiplicity-free
unitary irreducible representations}

In the following, we consider the problem of reduction of the unitary
irreducible $\overline{SL}(4,R)$ representations into the
corresponding $\overline{SL}(3,R)$ subgroup irreducible
representations. Due to complexity of these representations, we
confine in this work only to the multiplicity-free representations of
the $\overline{SL}(4,R)$ group, that themselves give rise, in the
reduction, to nonmultiplicity-free $\overline{SL}(3,R)$ representations.
Moreover, in order to maintain a parallelism between
infinite-component spinors and tensors, we shall consider here both
spinorial and tensorial representations. This parallelism is of
interest, for instance, when studying the infinite-component wave
equations of Regge-trajectory hadron recurrences \cite{R10}.  

Before proceeding further, let us concentrate on the parity properties of
the $\overline{SL}(4,R)$ generators, that can help considerably to
simplify the actual decomposition of considered representations. In the
$3+1$ notation, the $\overline{SL}(4,R)$ generators decompose
w.r.t. $\overline{SO}(3)$ into one quadrupole $T^{(2)}$, three vector
$J^{(1)}$, $N^{(1)}$, $K^{(1)}$, and one scalar $S^{(0)} \sim Q_{00}$
operator. 

The action of the parity (space inversion) operator $\cal P$ on the
$\overline{SL}(4,R)$ generators is as follows: 
\begin{eqnarray}
&&{\cal P}J{\cal P}^{-1} = +J, \quad {\cal P}T{\cal P}^{-1} = +T, \quad
{\cal P}S{\cal P}^{-1} = +S \nonumber \\ 
&&{\cal P}N{\cal P}^{-1} = -N, \quad {\cal P} K {\cal P}^{-1} = -K.
\end{eqnarray}
The operators $J$ and $T$ connect mutually the Hilbert space states
of the same parity, i.e. all the states of an $\overline{SL}(3,R)$
irreducible representation have the same parity. For example, the $J^P$,
i.e. the spin, parity content of the unitary irreducible
representation $D^{pr}_{\overline{SL}(3,R)}(0; \sigma_2 , \delta_2 )$ is
\begin{equation}
\{ J^P \} = \{ 0^+, 2^+, 4^+, \dots ; 2^+, 3^+, 4^+, \dots ; 4^+, 5^+,
6^+, \dots \}.
\end{equation}
The repeated action of the $N$ and $K$ operators results in the states of
alternating parities. For example, the $J^P$ content of a
finite-dimensional $D(\frac{7}{2},2)$ representation that is
$\overline{SO}(4)$  unitary, i.e. $\overline{SO}(1,3)$ nonunitary reads
\begin{equation}
\{ J^P \} = \{ \frac{3}{2}^+, \frac{5}{2}^-, \frac{7}{2}^+,
\frac{9}{2}^-, \frac{11}{2}^+ \} \quad \hbox{or}\quad \{ J^P \} =
\{ \frac{3}{2}^-, \frac{5}{2}^+, \frac{7}{2}^-, \frac{9}{2}^+,
\frac{11}{2}^- \}
\end{equation}

Let us consider at first the reduction of the simplest multiplicity-free
unitary irreducible representations of the $\overline{SL}(4,R)$ group,
i.e. the ladder ones $D^{(ladd)}_{\overline{SL}(4,R)}(0;e_2)$ and
$D^{(ladd)}_{\overline{SL}(4,R)}(\frac{1}{2};e_2)$. The
$\overline{SL}(4,R)$ $\supset$ $\overline{SO}(3)\otimes \overline{SO}(3)$
$\supset$ $\overline{SO}(3)$ decomposition of the
$D^{(ladd)}_{\overline{SL}(4,R)}(0;e_2)$ representation, i.e. the $\{
(j_1,j_2)\} \supset \{ J^P \}$ content reads
$$
\begin{array}{cccccc} \{(j_1,j_2)\} & = & \{(0,0), & (1,1), & (2,2), &
\dots\} \\ \cup &  & \cup & \cup & \cup & \\ J^P & = & 0^+ & 2^+ &
4^+ & \dots \\ & & & 1^- & 3^- & \dots \\ & & & 0^+ & 2^+ & \dots \\ & &
& & 1^- & \dots \\ & & & & 0^+ & \dots
\end{array} 
$$
The action of the noncompact operators $T$ and $S$ connects various
$\overline{SO}(3)$ states $J^P_{(j_1,j_2)}$ (where $(j_1,j_2)$ denotes the
"parent" $\overline{SO}(4)$ state of a given state $J$). The
irreducible actions of $T$ are
\begin{eqnarray*}
T: &&\{ 0^+_{(0,0)} \leftrightarrow 2^+_{(1,1)} \leftrightarrow
4^+_{(2,2)} \leftrightarrow \dots\},\quad \{ 1^-_{(1,1)} \leftrightarrow
3^-_{(2,2)} \leftrightarrow 5^-_{(3,3)} \leftrightarrow \dots\}, \\  
&&\{ 0^+_{(1,1)} \leftrightarrow 2^+_{(2,2)} \leftrightarrow 4^+_{(3,3)}
\leftrightarrow \dots\},\quad \{ 1^-_{(2,2)} \leftrightarrow 3^-_{(3,3)}
\leftrightarrow 5^-_{(4,4)} \leftrightarrow \dots\}, \dots
\end{eqnarray*}
The irreducible actions of $S$ are:
\begin{eqnarray*}
S: &&\{ 0^+_{(0,0)} \leftrightarrow 0^+_{(1,1)} \leftrightarrow
0^+_{(2,2)} \leftrightarrow \dots\},\quad \{ 1^-_{(1,1)} \leftrightarrow
1^-_{(2,2)} \leftrightarrow 1^-_{(3,3)} \leftrightarrow \dots\}, \\  
&&\{ 2^+_{(1,1)} \leftrightarrow 2^+_{(2,2)} \leftrightarrow 2^+_{(3,3)}
\leftrightarrow \dots\},\quad \{ 3^-_{(2,2)} \leftrightarrow 3^-_{(3,3)}
\leftrightarrow 3^-_{(4,4)} \leftrightarrow \dots\}, \dots
\end{eqnarray*}
Thus, we see that each $(j_1,j_2)=(j,j)\neq (0,0)$ is an "origin" of a
pair of $\overline{SL}(3,R)$ irreducible representations
$D^{ladd}_{\overline{SL}(3,R)}(0;\sigma_2)$ and
$D^{ladd}_{\overline{SL}(3,R)}(1;\sigma_2)$, while $(j_1,j_2) = (0,0)$ is
an "origin" of a single $\overline{SL}(3,R)$ irreducible representation,
$D^{ladd}_{\overline{SL}(3,R)}(0;\sigma_2)$. 
Symbolically, we write,
\begin{equation}
D^{ladd}_{\overline{SL}(4,R)}(0;e_2) \supset
D^{ladd}_{\overline{SL}(3,R)}(0;\sigma_2) \bigoplus_{j=1}^{\infty}
[D^{ladd}_{\overline{SL}(3,R)}(0;\sigma_2(j)) \oplus
D^{ladd}_{\overline{SL}(3,R)}(1;\sigma_2(j))] .
\end{equation}
The reduction of the $D^{ladd}_{\overline{SL}(4,R)}(\frac{1}{2};e_2)$
proceeds analogously.

Let us now consider the reduction of an $\overline{SL}(4,R)$ unitary
irreducible representation that has a nontrivial multiplicity. We
shall, for simplicity reasons, illustrate the method in the
$D^{pr}_{\overline{SL}(4,R)}(0,0;e_2)$ representation case. The
$\overline{SL}(4,R)$ $\supset$ $\overline{SO}(3)\otimes
\overline{SO}(3)$ $\supset$ $\overline{SO}(3)$ decomposition of the
$D^{(ladd)}_{\overline{SL}(4,R)}(0;e_2)$ representation, i.e. the $\{
(j_1,j_2)\} \supset \{ J^P \}$ content reads
$$
\begin{array}{cccccccccc} \{(j_1,j_2)\} & = & \{(0,0); & (2,0), & (1,1),
& (0,2); & (4,0), & (3,1), & (2,2),& \dots\} \\ \cup &  & \cup & \cup &
\cup & \cup & \cup & \cup & \cup & \\ J^P & = & 0^+ & 2^{\pm} & 2^+ &
2^{\pm} & 4^{\pm} & 4^{\pm} & 4^+ & \dots \\ & & & & 1^- & & & 3^{\mp} &
3^- & \dots \\ & & & & 0^+ & & & 2^{\pm} & 2^+ & \dots \\ & & & & & & & &
1^- & \dots \\ & & & & & & & & 0^+ & \dots
\end{array}
$$
The transitions: ${(j_1+j_2)}_{(j_1,j_2)}$ $\rightarrow$ ${(j_1+j_2\pm
2)}_{(j_1\pm 1,j_2\pm 1)}$ are due to the $T$ operator solely, and thus
these states have the same parity. The transitions
${(j_1+j_2)}_{(j_1,j_2)}$ $\rightarrow$ ${(j_1+j_2)}_{(j_1\pm 1,j_2\mp
1)}$ are due to the actions of the linear combinations $T\pm K$, and thus
the resulting states are not the eigenstates of the parity operator $\cal
P$. The states of definite parity are the symmetric and anti-symmetric
combinations of the corresponding states of the $(j_1,j_2)$ and
$(j_2,j_1)$ multiplets. For example, $4^+_{(3,1)+(1,3)} \equiv
4_{(3,1)}+4_{(1,3)}$ and $4^-_{(3,1)+(1,3)} \equiv
4_{(3,1)}-4_{(1,3)}$ are the eigenstates of positive and
negative parities, respectively. Owing to the nonvanishing matrix elements
of the $\overline{SL}(3,R)$ quadrupole operator $T$ between the
${(j_1+j_2)}_{(j_1,j_2)}$ and ${(j_1+j_2)}_{(j_1\pm 1,j_2\mp 1)}$ states,
we obtain in the reduction the $\overline{SL}(3,R)$ representations with
nontrivial multiplicity of $\overline{SO}(3)$ subrepresentations. We have
now all the information to regroup the $J^P$ states according to the
$\overline{SL}(3,R)$ irreducible representations. The lowest-lying
$D^{pr}_{\overline{SL}(4,R)}(0,0;e_2)$ states organize w.r.t. the $T$
operator action as follows:
$$
\{ 0^+_{(0,0)}, 2^+_{(1,1)}, 4^+_{(2,2)}, \dots ; 2^+_{(2,0)+(0,2)},
3^+_{(3,1)+(1,3)}, 4^+_{(3,1)+(1,3)} \dots ; 4^+_{(4,0)+(0,4)}, \dots \},
$$
\begin{eqnarray*}
\{ 0^+_{(1,1)}, 2^+_{(2,2)}, 4^+_{(3,3)}, \dots ; 2^+_{(3,1)+(1,3)},
3^+_{(4,2)+(2,4)}, 4^+_{(4,2)+(2,4)} \dots ; 4^+_{(5,1)+(1,5)}, \dots \},
\\
\{ 1^-_{(1,1)}, 3^-_{(2,2)}, 5^-_{(3,3)}, \dots ; 2^-_{(2,0)+(0,2)},
3^-_{(3,1)+(1,3)}, 4^-_{(3,1)+(1,3)} \dots ; 4^-_{(4,0)+(0,4)}, \dots \},
\end{eqnarray*}
\begin{eqnarray*}
\{ 0^+_{(2,2)}, 2^+_{(3,3)}, 4^+_{(4,4)}, \dots ; 2^+_{(4,2)+(2,4)},
3^+_{(5,3)+(3,5)}, 4^+_{(5,3)+(3,5)} \dots ; 4^+_{(6,2)+(2,6)}, \dots \},
\\
\{ 1^-_{(2,2)}, 3^-_{(3,3)}, 5^-_{(4,4)}, \dots ; 2^-_{(3,1)+(1,3)},
3^-_{(4,2)+(2,4)}, 4^-_{(4,2)+(2,4)} \dots ; 4^-_{(5,1)+(1,5)}, \dots \},
\end{eqnarray*}
$$
\dots
$$
It is seen from these expressions that there is a single
$\overline{SL}(3,R)$ irreducible representation,
$D^{pr}_{\overline{SL}(3,R)}(0;\sigma_2,\delta_2)$, "originating" from
the state $(j_1,j_2) = (0,0)$, while there is a pair of irreducible
representations, $D^{pr}_{\overline{SL}(3,R)}(0;\sigma_2,\delta_2)$ and
$D^{pr}_{\overline{SL}(3,R)}(1;\sigma_2,\delta_2)$ "originating" from
each set of states $\{(j_1,j_2)\ |\ j_1+j_2=2j,\ j=1,2,3,\dots \}$.
Symbolically, we write
\begin{eqnarray}
D^{pr}_{\overline{SL}(4,R)}(0,0;e_2) &\supset
&D^{pr}_{\overline{SL}(3,R)}(0;\sigma_2,\delta_2) \\
&&\bigoplus_{j=1}^{\infty}
[D^{pr}_{\overline{SL}(3,R)}(0;\sigma_2(j),\delta_2(j)) \oplus
D^{pr}_{\overline{SL}(3,R)}(1;\sigma_2(j),\delta_2(j))] .\nonumber
\end{eqnarray}
The reduction of all other nontrivial-multiplicity representations
proceeds analogously.

We list here the results of reductions of all multiplicity free
unitary irreducible representations of the $\overline{SL}(4,R)$ group
into the irreducible representations of its $\overline{SL}(3,R)$ subgroup.

\underline{\hbox{\it Principal Series:}}

\begin{eqnarray}
D^{pr}_{\overline{SL}(4,R)}(0,0;e_2) &\supset&
D^{pr}_{\overline{SL}(3,R)}(0_0;\sigma_2,\delta_2)
\\ &&\bigoplus_{j=1}^{\infty}
[D^{pr}_{\overline{SL}(3,R)}(0_0;\sigma_2(j),\delta_2(j)) \oplus
D^{pr}_{\overline{SL}(3,R)}(1_0;\sigma_2(j),\delta_2(j))], \nonumber \\
D^{pr}_{\overline{SL}(4,R)}(1,0;e_2) &\supset&
\bigoplus_{j=1}^{\infty}
[D^{pr}_{\overline{SL}(3,R)}(1_1;\sigma_2(j),\delta_2(j)) \oplus \bar
D^{pr}_{\overline{SL}(3,R)}(1_1;\sigma_2(j),\delta_2(j))] \nonumber
\end{eqnarray}

\underline{\hbox{\it Supplementary Series:}}

\begin{eqnarray}
D^{supp}_{\overline{SL}(4,R)}(0,0;e_1) &\supset
&D^{supp}_{\overline{SL}(3,R)}(0_0;\sigma_2,\delta_1) \\
&&\bigoplus_{j=1}^{\infty}
[D^{supp}_{\overline{SL}(3,R)}(0_0;\sigma_2(j),\delta_1(j)) \oplus
D^{supp}_{\overline{SL}(3,R)}(1_0;\sigma_2(j),\delta_1(j))] \nonumber
\end{eqnarray}

\underline{\hbox{\it Discrete Series:}}

\begin{eqnarray}
D^{disc}_{\overline{SL}(4,R)}(j_0,0) &\supset&
\bigoplus_{j=1}^{\infty}
D^{disc}_{\overline{SL}(3,R)}(j_0;\sigma_2(j),\delta_1(j)) \\
D^{disc}_{\overline{SL}(4,R)}(0,j_0) &\supset&
\bigoplus_{j=1}^{\infty}
D^{disc}_{\overline{SL}(3,R)}(j_0;\sigma_2(j),\delta_1(j)) \nonumber
\end{eqnarray}

\underline{\hbox{\it Ladder Series:}}

\begin{eqnarray}
D^{ladd}_{\overline{SL}(4,R)}(0;e_2) &\supset&
D^{ladd}_{\overline{SL}(3,R)}(0;\sigma_2) \\ 
&&\bigoplus_{j=1}^{\infty}
[D^{ladd}_{\overline{SL}(3,R)}(0;\sigma_2(j)) \oplus
D^{ladd}_{\overline{SL}(3,R)}(1;\sigma_2(j))] , \nonumber \\
D^{ladd}_{\overline{SL}(4,R)}(\frac{1}{2};e_2) &\supset&
\bigoplus_{j=1}^{\infty}
[D^{ladd}_{\overline{SL}(3,R)}(0;\sigma_2(j)) \oplus
D^{ladd}_{\overline{SL}(3,R)}(1;\sigma_2(j))] \nonumber
\end{eqnarray}

\section{$\overline{SL}(3,R)$ spinorial wave equation}

When embedding $\overline{SL}(3,R)$ into $\overline{SL}(4,R)$, there
are, as seen above, two (mutually contragradient) $\overline{SL}(3,R)$
vector candidates, i.e. $X \sim A = \frac{1}{2}(N+K)$ or $X \sim B =
\frac{1}{2}(N-K)$.  The explicit form of the $N$ operator (in the
spherical basis of the $\overline{SO}(4) = SU(2)\otimes SU(2)$ group)
is well known, while the embedding approach yields a closed
expressions for the $K$ operator as well. In particular, 
\begin{equation}
K_{\alpha} = (-)^{1-\alpha}i\sqrt{6} \left(\begin{array}{ccc}\ 1 & 1 & 1
\\ -\alpha & \beta & \gamma \end{array}\right) Z_{\beta \gamma} ,
\end{equation}
where $\alpha , \beta , \gamma = 0, \pm 1$, and thus its matrix
elements are given explicitly in terms of the $Z$ ones.

In the $\left|\begin{array}{c}J \\ M\end{array}\right>$ basis of the
$\overline{SO}(3) \subset \overline{SL}(3,R)$, one has,
\begin{eqnarray}
&&\left<\begin{array}{c}J' \\ M'\end{array}\right| K_{\alpha}
\left|\begin{array}{c}J \\ M\end{array}\right> \nonumber \\  
&&= i\sqrt{6} \sqrt{(2J'+1)(2J+1)(2j'_1+1)(2j_1+1)(2j'_2+1)(2j_2+1)}
\nonumber \\ 
&&\times\sum\Big[ (-)^{1-\alpha}(-)^{j'_1-m'_1}(-)^{j'_2-m'_2}
\left(\begin{array}{ccc}\ J' & j'_1 & j'_2 \\ -M' & m'_1 & m'_2
\end{array}\right) \left(\begin{array}{ccc}\ J & j_1 & j_2\\
-M & m_1 & m_2\end{array}\right) \nonumber \\ 
&&\left(\begin{array}{ccc}\ 1 & j'_1
& j'_2 \\ -\beta & m'_1 & m'_2\end{array}\right)
\left(\begin{array}{ccc}\ 1 & j_1 & j_2 \\ -\gamma & m_1 & m_2
\end{array}\right) \left(\begin{array}{ccc}\ 1 & 1 & 1 \\ -\alpha & \beta
& \gamma
\end{array}\right)\Big] \nonumber \\
&&<j'_1 j'_2 || Z || j_1 j_2 > .
\end{eqnarray}
The sum of the $3$-$j$ symbols in this expression is given in terms of the
$9$-$j$ symbol, and thus, we write
\begin{eqnarray}
\left<\begin{array}{c}J' \\ M'\end{array}\right| K_{\alpha}
\left|\begin{array}{c}J \\ M\end{array}\right>  &=&
i\sqrt{6}(-)^{J'-M'} \sqrt{(2J'+1)(2J+1)} \left(\begin{array}{ccc}\ J' &
1 & J \\ -M' & \alpha & M \end{array}\right) \nonumber \\ &&\times\left\{
\begin{array}{ccc} j'_1 & 1 & j_1 \\ j'_2 & 1 & j_2 \\ J' & 1 & J
\end{array} \right\} <j'_1 j'_2 || Z || j_1 j_2 > ,
\end{eqnarray}
where, $< j'_1 j'_2 || Z || j_1 j_2 >$ are the reduced matrix elements of
the operator $Z_{\alpha\beta}$.

Finally, we can write an $\overline{SL}(3,R)$ covariant wave equation in
the form
\begin{eqnarray}
&&(iX^{\mu}\partial_{\mu} - M) \Psi (x) = 0 , \\ 
&& \Psi\ \sim\ D^{disc}_{\overline{SL}(4,R)}(j_0,0),\
D^{disc}_{\overline{SL}(4,R)}(0,j_0) ,  \\
&&X^{\mu} = \frac{1}{2}(N^{\mu} + K^{\mu}) = \frac{1}{2}(J^{(1)\mu}
- J^{(2)\mu} + K^{\mu}) , 
\end{eqnarray}
where $\mu = 0, 1, 2$, and $K^0 = K_0,\ K^1 =
-\frac{1}{\sqrt{2}}(K_{+1} - K_{-1}),\ K^2 = \frac{1}{\sqrt{2}}(K_{+1} +
K_{-1})$. The matrix elements of all operators defining the
$\overline{SL}(3,R)$ vector operator $X^{\mu}$ in the infinite-component
representation of the field $\Psi (x)$ are explicitly constructed.

\section{Appendix A. $\overline{SL}(3,R)$ unitary irreducible
representations}

The unitary irreducible representations of the $\overline{SL}(3,R)$ group
\cite{R11} are defined in Hilbert spaces which are symmetric homogeneous
spaces over certain quotient subgroups $\cal K$ of its maximal compact
subgroup $\overline{SO}(3) \simeq SU(2)$. In other words, they are defined
in the spaces $L^{2}({\cal K})$ of square-integrable functions w.r.t. the
invariant measure $dk$ over ${\cal K}$, i.e.
$$
(f,g) = \int\int_{\cal K\otimes\cal K } f^*(k')\kappa (k',k'') g(k'') 
dk' dk'' .
$$
As a matter of fact, in order to account for nontrivial multiplicity of
the $\overline{SO}(3)$ subrepresentations, we work in the space of
functions over the diagonal subgroup $[{\cal K}_L\otimes{\cal K}_R]^d$
corresponding to the left and right group action, respectively. Thus,
there is another label, $K$, that accounts for nontrivial multiplicity.
In order to obtain all representations, one has to considered the
most general scalar product of the Hilbert space elements with, in
general, a nontrivial kernel $\kappa$. Furthermore, the irreducibility
requirements yield, in general, certain relationships between the
representation labels, the corresponding labels of the maximal compact
subgroup and the matrix elements of $\kappa$.

When ${\cal K} = SU(2)$, the representation space basis is $\left|
\begin{array}{c} J \\ K\ M \end{array} \right>$. The compact generators
matrix elements are the well known ones,
\begin{eqnarray*}
J_0\left|\begin{array}{c} J \\ K\ M\end{array}\right> &=& M
\left|\begin{array}{c} J\\ K\ M \end{array}\right> , \\
J_{\pm}\left|\begin{array}{c}J\\ K\ M\end{array}\right>  &=&
\sqrt{J(J+1)-M(M\pm 1)} \left|\begin{array}{c}J\\K\ M\pm 1
\end{array}\right>
\end{eqnarray*}
while the matrix elements of the noncompact generators are given by the
following expression:
\begin{eqnarray*}
&&\left<\begin{array}{c} J' \\ K'\ M'\end{array}\right| T_{\rho}
\left|\begin{array}{c} J \\ K\ M\end{array}\right> \\ &&= -i(-)^{J'-K'}
(-)^{J'-M'} \sqrt{(2J'+1)(2J+1))} \left(\begin{array}{ccc}\ J' & 2 & J \\
-M' & \rho & M \end{array}\right)\\ &&\times\Big[ \left(
\sqrt{\frac{2}{3}} (\sigma_1+i\sigma_2-\frac{1}{\sqrt{6}}
(J'(J'+1)-J(J+1)) \right) \left(\begin{array}{ccc}\ J' & 2 & J \\ -K' & 0
& K \end{array}\right) \\ &&- (\delta_1+1+i\delta_2)
\left(\begin{array}{ccc} \ J' & 2 & J \\ -K' & 2 & K \end{array}\right) -
(\delta_1+1+i\delta_2) \left(\begin{array}{ccc} \ J' &\ 2 & J \\ -K' & -2
& K \end{array}\right)\Big]
\end{eqnarray*}
where, $\sigma' = \sigma_1 + i\sigma_2$, and $\delta' = \delta_1 +
\delta_2$ ($\sigma_1$, $\sigma_2$, $\delta_1$, $\delta_2$ $\in$ $R$) are
the $\overline{SL}(3,R)$ representation label. The $\{ J \}$ content,
i.e. $\overline{SO}(3)$ subrepresentations, follow the two general rules:
(i) $\Delta K = 0, \pm 1$, and (ii) for each $K \neq 0$, $J = K, K+1,
K+2, \dots$, while for $K=0$, either $J = 0, 2, 4, \dots$ or $J = 1, 3,
5,\dots$.

When ${\cal K} = SU(2)/U(1)$, the representation space basis is $\left|
\begin{array}{c} J \\ M \end{array} \right>$. The compact generators
matrix elements are the well known ones,
\begin{eqnarray*}
J_0\left|\begin{array}{c} J \\ M\end{array}\right> &=& M
\left|\begin{array}{c} J\\ M \end{array}\right> , \\
J_{\pm}\left|\begin{array}{c}J\\ M\end{array}\right>  &=&
\sqrt{J(J+1)-M(M\pm 1)} \left|\begin{array}{c}J\\ M\pm 1
\end{array}\right>
\end{eqnarray*}
while the matrix elements of the noncompact generators are given by the
following expression:
\begin{eqnarray*}
&&\left<\begin{array}{c} J' \\ M'\end{array}\right| T_{\rho}
\left|\begin{array}{c} J \\ M\end{array}\right> \\ &&= -i(-)^{J'}
(-)^{J'-M'} \sqrt{(2J'+1)(2J+1)} \left(\begin{array}{ccc}\ J' & 2 & J \\
-M' & \rho & M \end{array}\right)\\ &&\times \left( \sqrt{\frac{2}{3}}
(\sigma_1+i\sigma_2-\frac{1}{\sqrt{6}} (J'(J'+1)-J(J+1)) \right)
\end{eqnarray*}
where, $\sigma_1, \sigma_2 \in R$ are the $\overline{SL}(3,R)$
representation labels. The $3$-$j$ symbol $\left( \matrix{J' & 2 & J \cr 0
& 0 & 0 }\right)$, with half-integer entries is to be evaluated by taking
the explicit expression for the integer case and continuing it to the
half-integer one. The $\{ J \}$ content, i.e. $\overline{SO}(3)$
subrepresentations, is given by the rule $\Delta J = 0, \pm 2$.

There are, besides the trivial representation, four series of unitary
irreducible representations of the $\overline{SL}(3,R)$ group, that are
characterized by the representation label, the minimal $\underline{J}$
(and when necessary the minimal $\underline{K} \ge 0$) values, and they
are defined in Hilbert spaces with the basis vectors corresponding to
certain irreducible lattices in the $J-|K|$ plane, and the scalar
products are given in terms of the kernel $\kappa$, i.e.
$D_{\overline{SL}(3,R)}(\underline{J}_{\underline{K}};\sigma ,\delta )$

{\it Principal Series}.

\begin{eqnarray*}
&&D^{pr}_{\overline{SL}(3,R)}(0_0;\sigma_2 ,\delta_2 ),\quad \{ J\} = \{
0, 2, 4, \dots ; 2, 3, 4, \dots ; 4,5,6,\dots ;\dots \} \\
&&D^{pr}_{\overline{SL}(3,R)}(1_0;\sigma_2 ,\delta_2 ),\quad \{ J\} = \{
1, 3, 5, \dots ; 2, 3, 4, \dots ; 4,5,6,\dots ;\dots \} \\
&&D^{pr}_{\overline{SL}(3,R)}(1_1;\sigma_2 ,\delta_2 ),\quad \{ J\} = \{
1, 2, 3, \dots ; 3, 4, 5, \dots ; 5,6,7, \dots ;\dots \} \\
&&D^{pr}_{\overline{SL}(3,R)}(\frac{1}{2};\sigma_2 ,\delta_2 ),\quad \{
J\} = \{ \frac{1}{2}, \frac{3}{2}, \frac{5}{2}, \dots ; \frac{3}{2},
\frac{5}{2}, \frac{7}{2}, \dots ;\frac{5}{2}, \frac{7}{2}, \frac{9}{2},
\dots  ;\dots \},
\end{eqnarray*}
where $\sigma_2 ,\delta_2 \in R$. They are defined in the Hilbert spaces
$H(SU(2),\kappa )$, where $\kappa =\sqrt{2J+1}$, $\forall J$.

{\it Supplementary Series}.

\begin{eqnarray*}
&&D^{supp}_{\overline{SL}(3,R)}(0_0;\sigma_2 ,\delta_1 ),\quad \{ J\} =
\{ 0, 2, 4, \dots ; 2, 3, 4, \dots ; 4,5,6,\dots ;\dots \} \\
&&D^{supp}_{\overline{SL}(3,R)}(1_0;\sigma_2 ,\delta_1 ),\quad \{ J\} =
\{ 1, 3, 5, \dots ; 2, 3, 4, \dots ; 4,5,6,\dots ;\dots \} \\
&&D^{supp}_{\overline{SL}(3,R)}(\frac{1}{2};\sigma_2 ,\delta_1 ),\quad \{
J\} = \{ \frac{1}{2}, \frac{3}{2}, \frac{5}{2}, \dots ; \frac{3}{2},
\frac{5}{2}, \frac{7}{2}, \dots ;\frac{5}{2}, \frac{7}{2}, \frac{9}{2},
\dots  ;\dots \},
\end{eqnarray*}
where $\sigma_2 \in R$, while $|\delta_1| < 1$ for integer $J$
($\underline{K}=0$ only), and $|\delta_1| < \frac{1}{2}$ for half integer
$J$ ($\underline{K}=\frac{1}{2}$). They are defined in the Hilbert spaces
$H(SU(2),\kappa )$, and the $\kappa$ matrix elements are $\kappa (J;K) =
\sqrt{\frac{2J+1}{2\underline{J}+1}} \frac{\Gamma
(\frac{1}{2}(K+1-\delta_1)) \Gamma
(\frac{1}{2}(\underline{K}+1+\delta_1))}{\Gamma
(\frac{1}{2}(K+1+\delta_1))\Gamma
(\frac{1}{2}(\underline{K}+1-\delta_1))}$.

{\it Discrete Series}.

\begin{eqnarray*}
D^{disc}_{\overline{SL}(3,R)}(\underline{J}; \sigma_2 ,\delta_1 ), &&\{
J\} = \{ \underline{J}, \underline{J}+1, \underline{J}+2, \dots ;
\underline{J}+2, \underline{J}+3, \underline{J}+4, \dots ; \\
&&\underline{J}+4,\underline{J}+5,\underline{J}+6, \dots ; \dots \}
\end{eqnarray*}
where $\underline{J} = \underline{K} = \frac{3}{2}, 2, \frac{5}{2}, 3,
\dots $, and $\delta_1 = 1 - \underline{J}$. They are defined in the
Hilbert spaces $H(SU(2),\kappa )$, and 
%the $\kappa$ matrix elements are
$\kappa (J;K)$ $=$ $\sqrt{\frac{2J+1}{2\underline{J}+1}} \frac{\Gamma
(\frac{1}{2}(K+\underline{K}))}{\Gamma (\frac{1}{2}(K-\underline{K}))
\Gamma (\underline{K})}$.

{\it Ladder Series}.

\begin{eqnarray*}
&&D^{ladd}_{\overline{SL}(3,R)}(0; \sigma_2 ),\quad \{ J\} = \{ 0, 2, 4,
\dots \} \\ &&D^{ladd}_{\overline{SL}(3,R)}(0; \sigma_2 ),\quad \{ J\} =
\{ 1, 3, 5, \dots \} \\
&&D^{ladd}_{\overline{SL}(3,R)}(\frac{1}{2}),\quad\quad \{ J\} = \{
\frac{1}{2}, \frac{5}{2}, \frac{9}{2}, \dots \}
\end{eqnarray*}
where $\sigma_2 \in R$. They are defined in the Hilbert spaces
$H(SU(2)/U(1),\kappa )$, and the $\kappa$ matrix elements are $\kappa
(J;K)= \sqrt{2J+1}$, $\forall J$. Note that, owing to the unitarity
requirement, there is a unique spinorial unitary irreducible
representation $D^{ladd}_{\overline{SL}(3,R)}(\frac{1}{2})$ corresponding
to $\sigma_2=0$.

\section{Appendix B. $\overline{SL}(4,R)$ multiplicity-free unitary
irreducible representations}

The unitary irreducible representations of the $\overline{SL}(4,R)$
group, that are  multi\-plicity-free w.r.t. its maximal compact subgroup
$\overline{SO}(4) \simeq SU(2)\otimes SU(2)$ \cite{R12}, are defined in
Hilbert spaces which are symmetric homogeneous spaces over certain quotient   
subgroups $\cal K$ of the maximal compact subgroup. In other words, they
are defined in the spaces $L^{2}(\cal K)$ of square-integrable functions
w.r.t. the invariant measure $dk$ over ${\cal K}$, i.e. 
$$
(f,g) =
\int\int_{{\cal K}\otimes{\cal K}}f^*(k')\kappa (k',k'')g(k'')dk'dk''.
$$ 
In order to obtain all these representations, one has to considered the
most general scalar product of the Hilbert space elements with, in
general, a nontrivial kernel $\kappa$. Furthermore, the representation
irreducibility requirement implies that the $K$ representation
eigenvector labels, $j_1$, $j_2$, which define a basis of the
$\overline{SL}(4,R)$ representation Hilbert space, are constrained to
belong to certain invariant lattices $L$ of the $(j_1,j_2)$ plane of
points. Therefore, we denote the unitary irreducible representation
Hilbert spaces symbolically by $H({\cal K}, \kappa ,L)$.

When ${\cal K} = [SU(2)/U(1)]\otimes [SU(2)/U(1)]$, the representation
space basis is $\left| \begin{array}{cc} j_1 & j_2 \\ m_1 & m_2
\end{array} \right>$. The compact generators matrix elements are the well
known ones,
\begin{eqnarray*}
J_0^{(1)} \left|\begin{array}{cc} j_1 & j_2\\ m_1 & m_2
\end{array}\right> &=& m_1 \left|\begin{array}{cc}j_1 & j_2\\
m_1 & m_2 \end{array}\right> , \\
J_0^{(2)}\left|\begin{array}{cc}j_1 & j_2\\ m_1 & m_2
\end{array}\right> &=& m_2 \left|\begin{array}{cc}j_1 & j_2\\
m_1 & m_2 \end{array}\right> , \\
J_{\pm}^{(1)}\left|\begin{array}{cc}j_1 & j_2\\ m_1 & m_2
\end{array}\right>  &=& \sqrt{j_1(j_1 + 1) - m_1(m_1 \pm 1)}
\left|\begin{array}{cc}j_1 & j_2\\ m_1\pm 1 & m_2
\end{array}\right> , \\
J_{\pm}^{(2)} \left|\begin{array}{cc}j_1 & j_2 \\ m_1 & m_2
\end{array}\right> &=& \sqrt{j_2(j_2 + 1) - m_2(m_2 \pm 1)}
\left|\begin{array}{cc}j_1 & j_2 \\ m_1 & m_2 \pm 1
\end{array}\right>,
\end{eqnarray*}
while the matrix elements of the noncompact generators are given
by the following expression:
\begin{eqnarray*}
&&\left< \begin{array}{cc} j'_1 & j'_2 \\ m'_1 & m'_2
\end{array} \right|
Z_{\alpha\beta} \left|\begin{array}{cc} j_1 & j_2 \\ m_1 & m_2
\end{array}\right> \\
&&= -i(-)^{j'_1 - m'_1}(-)^{j'_2 - m'_2}(-)^{j'_1 + j'_2}
\sqrt{(2j'_1+1)(2j'_2+1)(2j_1+1)(2j_2+1)} \\ &&\times (e -
\frac{1}{2}[j'_1(j'_1+1)-j_1(j_1+1) + j'_2(j'_2+1)-j_2(j_2+1)])\\
&&\times \left(\begin{array}{ccc} \ j'_1 & 1 & j_1 \\ -m'_1 & \alpha &
m_1\end{array}\right) \left(\begin{array}{ccc}\ j'_2 & 1 & j_2 \\ -m'_2 &
\beta & m_2 \end{array}\right) \left(\begin{array}{ccc} j'_1 & 1 & j_1 \\
0 & 0 & 0
\end{array}\right) \left(\begin{array}{ccc} j'_2 & 1 & j_2 \\ 0 &
0 & 0 \end{array}\right) ,
\end{eqnarray*}
where, $e = e_1 + ie_2$, is the $\overline{SL}(4,R)$ representation
label. The $3$-$j$ symbol $\left( \matrix{j^{\prime} & 1 & j \cr 0 & 0 & 0
}\right)$, with half-integer entries is to be evaluated by taking the
explicit expression for the integer case and continuing it to the
half-integer one. There are, in this case, eight invariant lattices of
points in the space of the $SU(2)\times SU(2)$ representation labels, $L
= \{(j_1, j_2)\}$,  $j_1, j_2 = 0, \frac{1}{2}, 1, \frac{3}{2}, \dots $,
that are characterized by the conditions
$$
(j_1 + j_2) - (j_{01} + j_{02}) \equiv 0 (mod 2),\quad (j_1 -
j_2)- (j_{01} - j_{02}) \equiv 0 (mod 2),
$$
where $(j_{01}, j_{02})$ are the "minimal" $(j_1, j_2)$ values.
This is due to the noncompact generators action allowing only for
$(j'_1, j'_2) = (j_1\pm 1, j_2\pm 1)$. We define these invariant
lattices by their minimal $(j_1, j_2)$ values, i.e.
$$
\begin{array}{llll} L(0,0), & L(\frac{1}{2},\frac{1}{2}),
& L(1,0)=L(0,1), & L(\frac{3}{2}, \frac{1}{2})=L(\frac{1}{2},
\frac{3}{2}, \\ L(\frac{1}{2},0), & L(0, \frac{1}{2}), &
L(\frac{3}{2},0), & L(0,\frac{3}{2}).
\end{array}
$$

When ${\cal K} = [SU(2)\otimes SU(2)]/SU(2)$, one has $j_1 = j_2 \equiv
j$, and the representation space basis is $\left|
\begin{array}{c} j \\ m_1 \ m_2 \end{array} \right>$. The compact
generators matrix elements are
\begin{eqnarray*}
J_0^{(1)} \left|\begin{array}{c} j \\ m_1 \ m_2 \end{array}\right>
&=& m_1\left|\begin{array}{c}j \\ m_1 \ m_2 \end{array}\right> ,\\
J_0^{(2)} \left|\begin{array}{c}j \\ m_1 \ m_2
\end{array}\right> &=& m_2 \left|\begin{array}{c}j \\ m_1 \ m_2
\end{array}\right> ,\\
J_{\pm}^{(1)} \left|\begin{array}{c}j \\ m_1 \ m_2
\end{array}\right>  &=& \sqrt{j(j + 1) - m_1(m_1 \pm 1)}
\left|\begin{array}{c} j \\ m_1\pm 1 \ m_2 \end{array}\right> , \\
J_{\pm}^{(2)} \left|\begin{array}{c}j \\ m_1 \ m_2
\end{array}\right> &=& \sqrt{j(j + 1) - m_2(m_2 \pm 1)}
\left|\begin{array}{c}j \\ m_1 \ m_2 \pm 1 \end{array}\right>,
\end{eqnarray*}
while the matrix elements of the noncompact generators are given
by the following expression:
\begin{eqnarray*}
&&\left< \begin{array}{c} j' \\ m'_1\ m'_2 \end{array} \right|
Z_{\alpha\beta} \left| \begin{array}{c} j \\ m_1\ m_2\end{array} \right>
\\ &&= -i (-)^{j' - m'_1}(-)^{j' - m'_2}\sqrt{(2j' + 1)(2j+1)} (e -
\frac{1}{2} [j'(j' + 1) - j(j+1)]) \\ &&\times \left(
\begin{array}{ccc}\ j' & 1 & j \\ -m'_1 & \alpha & m_1 \end{array}
\right) \left( \begin{array}{ccc}\ j' & 1 & j \\ -m'_2 & \beta &
m_2 \end{array} \right) ,
\end{eqnarray*}
where $e = e_1 + ie_2$ is the $\overline{SL}(4,R)$ representation
label. There are, in this case, two invariant lattices:
\begin{eqnarray*}
L(0)&=& \{ (0, 0), (1, 1), \dots \} \\ L(\frac{1}{2}) &=& \{
(\frac{1}{2}, \frac{1}{2}), (\frac{3}{2}, \frac{3}{2}), \dots \}.
\end{eqnarray*}

There are, besides the trivial representation, four series of
multiplicity-free unitary irreducible representations of the
$\overline{SL}(4,R)$ group, that are characterized by the representation
label, the minimal $(j_1,j_2)$ values, and they are defined in Hilbert
spaces with the basis vectors corresponding to certain irreducible
lattices in the $j_1-j_2$ plane, and the scalar products given in terms
of the kernel $\kappa$.

{\it Principal Series}.
$$
D^{pr}_{\overline{SL}(4,R)}(0,0;e_2),\quad
D^{pr}_{\overline{SL}(4,R)}(1,0;e_2),\quad e_1=0,\ e_2\in R.
$$
They are defined in the Hilbert spaces $H([SU(2)/U(1)]\otimes
[SU(2)/U(1)],\kappa ,L)$, where $\kappa (j_{1},j_{2})=1, \forall
j_{1},j_{2}$, and the irreducible lattices are, respectively,
$L(0,0)$ and $L(1,0)$.

{\it Supplementary Series}.
$$
D^{supp}_{\overline{SL}(4,R)}(0,0;e_1),\quad 0 < |e_1| < 1,\
e_2=0.
$$
They are defined in the Hilbert spaces
$H([SU(2)/U(1)]\otimes [SU(2)/U(1)],\kappa ,L)$, where $\kappa
(j_1,j_2)$ is nontrivial and given by
$$
\kappa (j_1,j_2) = \frac{\Gamma (j_1+j_2+e_1+1)\Gamma (1-e_1)
\Gamma (|j_1-j_2|+e_1+2)\Gamma (2-e_1)}{\Gamma
(j_1+j_2-e_1+1)\Gamma (1+e_1) \Gamma (|j_1-j_2|-e_1+2)\Gamma
(2+e_1)},
$$
and the irreducible lattice is $L(0,0)$.

{\it Discrete Series}.
$$
D^{disc}_{\overline{SL}(4,R)}(j_0,0),\quad
D^{disc}_{\overline{SL}(4,R)}(0,j_0),\ e_1=1-j_0,\
j_0=\frac{1}{2}, 1, \frac{3}{2}, \dots ,\  e_2=0.
$$
They are defined in the Hilbert spaces $H([SU(2)/U(1)]\otimes
[SU(2)/U(1)],\kappa ,L)$, where $\kappa (j_{1},j_{2})$ is
nontrivial and given by
$$
\kappa (j_1,j_2) = \frac{\Gamma (j_1+j_2+e_1+1)\Gamma
(|j_1-j_2|+e_1+2)}{\Gamma (j_1+j_2-e_1+1)\Gamma
(|j_1-j_2|-e_1+2)},
$$
and the irreducible lattices are, respectively, $L(j_0,0 |
j_1-j_2\geq j_0)$ $\subset$ $L(0,0)$, $L(\frac{1}{2},0)$,
$L(1,0)$, $L(\frac{3}{2},0)$ and $L(0,j_0 | j_2-j_1\geq j_0)$
$\subset L(0,0)$, $L(0,\frac{1}{2})$, $L(0,1)$,
$L(0,\frac{3}{2})$.

{\it Ladder Series}.

\begin{eqnarray*}
D^{ladd}_{\overline{SL}(4,R)}(0; e_2), \quad
D^{ladd}_{\overline{SL}(4,R)}(\frac{1}{2};e_2),\quad e_1=0, e_2
\in R.
\end{eqnarray*}
They are defined in the Hilbert spaces $H([SU(2)\otimes
SU(2)]/SU(2),\kappa ,L)$, where $\kappa (j,j)=1, \forall j$, and
the irreducible lattices are, respectively, $L(0)$ and
$L(\frac{1}{2})$.

\section*{Acknowledgments}

This work was supported in part by MSE, Belgrade, Project-101486.

\end{document}